\documentclass[journal=jpclcd,manuscript=letter]{achemso}
\setkeys{acs}{usetitle=false}
\pdfoutput=1
\usepackage[version=3]{mhchem} 
\usepackage{natmove}
\usepackage{multirow}
\author{Caterina Cocchi}
\email{caterina.cocchi@unimore.it}
\affiliation{Centro S3, CNR-Istituto Nanoscienze, I-41125 Modena, Italy}
\alsoaffiliation{Dipartimento di Fisica, Universit\`a di Modena e Reggio Emilia, I-41125 Modena, Italy}
\author{Deborah Prezzi}
\affiliation{Centro S3, CNR-Istituto Nanoscienze, I-41125 Modena, Italy}
\author{Alice Ruini}
\affiliation{Centro S3, CNR-Istituto Nanoscienze, I-41125 Modena, Italy}
\alsoaffiliation{Dipartimento di Fisica, Universit\`a di Modena e Reggio Emilia, I-41125 Modena, Italy}
\author{Marilia J. Caldas}
\affiliation{Instituto de F{\'\i}sica, Universidade de S\~ao Paulo, 05508-900 S\~ao Paulo, SP, Brazil}
\author{Elisa Molinari}
\affiliation{Centro S3, CNR-Istituto Nanoscienze, I-41125 Modena, Italy}
\alsoaffiliation{Dipartimento di Fisica, Universit\`a di Modena e Reggio Emilia, I-41125 Modena, Italy}
\title{Optical properties and charge-transfer excitations in edge-functionalized all-graphene nanojunctions}
\begin{document}
\begin{abstract}
We investigate the optical properties of edge-functionalized graphene nanosystems, focusing on the formation of junctions and charge transfer excitons.
We consider a class of graphene structures which combine the main electronic features of graphene with the wide tunability of large polycyclic aromatic hydrocarbons.
By investigating prototypical ribbon-like systems, we show that, upon convenient choice of functional groups, low energy excitations with remarkable \textit{charge transfer character} and \textit{large oscillator strength} are obtained.
These properties can be further modulated through an appropriate width variation, thus spanning a wide range in the low-energy region of the UV-Vis spectra.
Our results are relevant in view of designing all-graphene optoelectronic nanodevices, which take advantage of the versatility of molecular functionalization, together with the stability and the electronic properties of graphene nanostructures.
\end{abstract}


\newpage
Research on the graphene chemistry \cite{loh+10jmc} has been stimulated
in the last few years by an increasing demand for both large scale production of 
graphene samples and atomic control on its nanostructures.
The electronic features of graphene, related to its $\pi$-electron network, make this novel material analogous to large polycyclic aromatic hydrocarbons (PAHs) \cite{tyut+98jpcb,wats+01cr} and allow to exploit a wide range of well-known functionalizations, reactions and preparation techniques, also useful for technological applications which need the blending of graphene with organic and inorganic compounds.
One of the main advantages of these chemical approaches is related to the possibility to obtain graphene nanostructures, tailoring their size, shape and edge termination on the atomic scale \cite{wu+07cr,cai+10nat}.
The control of these parameters can be exploited to tune the electronic \cite{qian+08jacs,wang+09sci,li+10jacs,shah+11} and optical \cite{prez+08prb,prez+07pssb,prez+11tobe,yang+07nl,zhu-su10jpcc} properties of graphene nanostructures.
\begin{figure}%
\centering
\includegraphics[width=.3\textwidth]{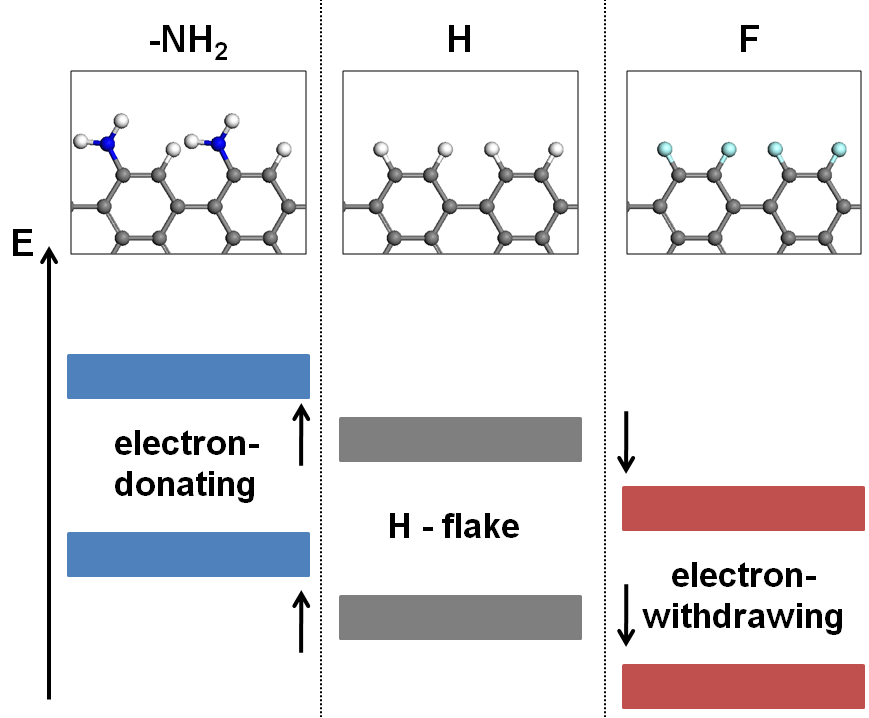}%
\caption{Scheme of the energy offset between edge covalently functionalized graphene nanoflakes, with respect to a hydrogenated flake (H-flake) taken as a reference. Edge functional groups with remarkable electron-donating (-withdrawing) character, such as \mbox{-\ce{NH2}} (F) induce an almost rigid upshift (downshift) of the gap region. }
\label{fig1}
\end{figure}

In this direction, we recently demonstrated \cite{cocc+11jpcc} that edge covalent functionalization represents a viable method to tune the ionization potential (IP) of graphene nanoflakes (GNFs), preserving the overall characteristics of the $\pi$-conjugation. 
Furthermore, as illustrated in \ref{fig1}, a corresponding shift of the electron affinity (EA) moves the \textit{energy gap region}  of the GNF in an almost rigid fashion according to the edge termination. 
With respect to a reference hydrogenated flake (see \ref{fig1}), an upshift comes from electron-donating and a downshift from electron-withdrawing functional groups. 
The effectiveness of this mechanism allows us to form all-graphene \textit{type II} nanojunctions (GNJs) which can be built such that occupied and virtual molecular orbitals (MOs) close to the gap region localize on opposite sides. 
This opens the possibility of obtaining charge transfer excitons, which are relevant for many optoelectronic applications.

Here we study the optical properties of such edge-functionalized GNJs, with particular attention to the conditions for the formation of charge transfer excitons. 
We start characterizing the UV-Vis spectra of hydrogenated and functionalized GNFs, as obtained by considering atomic fluorine (F), methyl-ketone (\mbox{-\ce{COCH3}}) and amino (\mbox{-\ce{NH2}}) groups.
We then focus on few prototypical nanojunctions, obtained by varying in turn edge functionalization and junction width. 
We show that upon appropriate modulation of these parameters, GNJs can be engineered in order to get large oscillator strength and charge transfer character for the first excitation.
Moreover, following the energy gap modulation with respect to the width, the energy of the first peak in the UV-Vis spectra can be tuned in order to span a wide range ($\sim$ 1.5 eV) in the low energy region.

The UV-Vis spectra are computed within the framework of the semi-empirical 
Hartree-Fock (HF) based method ZINDO \cite{ridl-zern73tca}, through the
configuration interaction (CI) procedure, including single excitations only (ZINDO/CIS). 
This method is known to provide reliable results for aromatic molecules 
\cite{cald+01apl,wang+04jacs}.
All calculations are performed starting from optimized geometries (0.4 
$\text{kcal} \cdot \text{mol}^{-1}$/\AA{} force threshold) and mean-field ground 
state properties calculated with AM1 \cite{dewa+85jacs,MS-note}.

%
\begin{figure}%
\centering
\includegraphics[width=.4\textwidth]{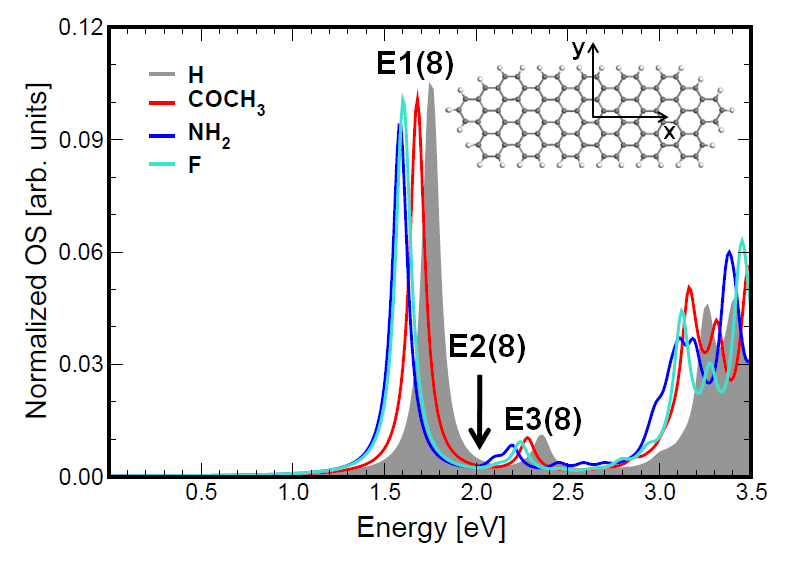}%
\caption{UV-Vis spectra of functionalized graphene
flakes (empty solid curves) and of the reference hydrogenated flake
(shaded gray area). The curves represent oscillator strength (OS),
normalized with respect to the number of C rings,
\textit{vs.} excitation energy and are convoluted with a Lorentzian
function, with a broadening of 0.05 eV.
The structure of the hydrogenated (H-) flake and the reference axes are also shown.} 
\label{fig2}
\end{figure}
We first discuss the optical properties of homogeneously functionalized graphene flakes in comparison with those of their hydrogenated counterpart (the structure is shown in \ref{fig2}). 
These systems are characterized by \textit{armchair}-shaped edges and width parameter $N=8$, where $N$ indicates the number of dimer lines along the \textit{zigzag} direction, borrowing the standard  notation used for quasi-1D armchair graphene nanoribbons (AGNRs) \cite{son+06prl}.
We have already shown \cite{cocc+11jpcc} that the class of graphene flakes considered here follows the width-modulated behavior of the energy gap ($E_G$ equal to the difference between IP and EA) typical of AGNRs, also upon edge functionalization.
In particular, while the band gap of AGNRs scales inversely with respect to the ribbon width \textit{w} \cite{son+06prl,baro+06nl}, three $\sim 1/w$ scaling laws have been identified depending on the \textit{N} parameter, expressed as $N=3p+m$ ($p$ is a positive integer and $m=0,1,2$).
The smallest gap pertains to the $N=3p+2$ family. 
The flakes considered here are about 3 nm long (\textit{x} direction) and their 
borders along the width (\textit{y} direction) are shaped in order to minimize 
their influence on molecular orbitals (MOs) and to prevent the presence of 
localized magnetic states \cite{hod+08prb}.
The functionalizations treated here comprise both \mbox{-\ce{NH2}} groups, which induce an upshift of the gap region with respect to the hydrogenated counterpart, and \mbox{-\ce{COCH3}} groups and F atoms, which on the contrary downshift the gap region, according to \ref{fig1}. 
The synthesis of fluorine substituted graphene molecules \cite{zhan+05ol,kiku+07ol} as well as functionalized with \mbox{-\ce{NH2}} \cite{wang+09sci} and ketone-based \cite{qian+08jacs,li+10jacs} groups is well known and has been experimentally demonstrated.
\ce{COCH3}- and \ce{NH2}-functionalized flakes are here obtained by replacing each second H-terminating atom along both edges with a covalently bonded group, while atomic fluorine substitutes each H edge atom in the case of F-flake (see upper panel of \ref{fig1}).
Both functionalized flakes and junctions display distortions due to steric effects induced by edge terminating groups \cite{dist-note}.

By inspecting the UV-Vis spectra of the flakes shown in \ref{fig2}, we notice that 
the optical properties of the systems are not significantly changed upon 
functionalization, at least in the low energy window. 
Only a slight red shift is noticed, on account of the reduced energy gap with 
respect to the hydrogenated flake: this effect is maximized in case of \ce{NH2}- 
and F-functionalization, which show almost equal electrical and optical gaps. 
The first excitation, labelled as E1(8) in \ref{fig2},
is optically active in all cases and dominated by a HOMO
$\rightarrow$ LUMO transition. The second peak in the spectra [E3(8)] is 
generated by the third excitation and is mainly due to \mbox{HOMO-1} 
$\rightarrow$ LUMO and HOMO $\rightarrow$ \mbox{LUMO+1} transitions. 
Although these transitions are allowed by symmetry, the overlap
between the involved orbitals is small and thus the oscillator
strength for E3(8) is reduced of about one order of magnitude with respect to E1(8). 
Between the two we find another excitation [E2(8)],
which is very weak but not forbidden by symmetry, thanks to a small
but non vanishing contribution of HOMO $\rightarrow$ \mbox{LUMO+1}
transition. Further details on the composition of the excitations
and on the symmetry of the involved MOs are reported in the
Supporting Information.

\begin{figure}
\centering
\includegraphics[width=.48\textwidth]{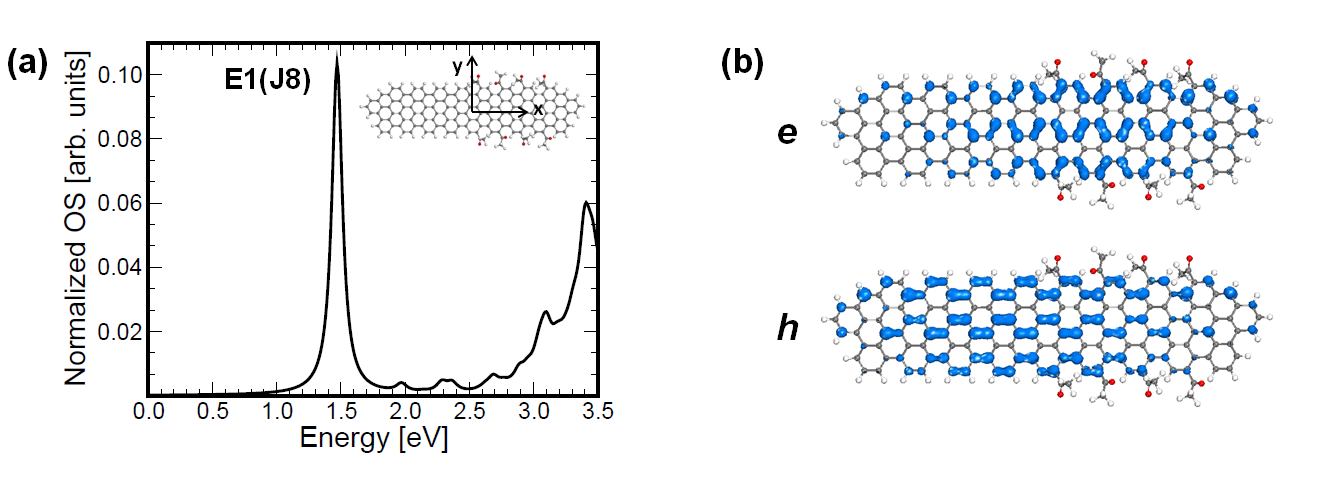}%
\caption{(a) UV-Vis spectrum of $N=8$ H//\ce{COCH3} nanojunction. The curve represents the oscillator strength (OS), normalized with respect to the number of C rings, \textit{vs.} excitation energy and is convoluted with a Lorentzian function, of broadening 0.05 eV.
The structure of the junction is also shown; (b) Electron (\textit{e}) and hole (\textit{h}) probability density for E1(J8) excitation of  H//\ce{COCH3} junction, after \ref{rhoe} and \ref{rhoh}.
}
\label{fig3}
\end{figure}
We next consider the optical properties of the nanojunction obtained
by interfacing H-termination and \ce{COCH3}-functionalization: this
system has the same width of the GNFs discussed above (i.e. $N=8$)
while its length ($\sim$ 4 nm) is tailored in such a way that each
side of the junction correctly accounts for the properties of the
corresponding homogeneously functionalized system. By inspecting
the UV-Vis spectrum of this junction, shown in \ref{fig3}(a), we
notice that the shape resembles that of isolated GNFs. However,
the character
of the first excitation, labelled as E1(J8) in \ref{fig3}(a),
changes dramatically. 
In fact, HOMO and LUMO of the H//\ce{COCH3} GNJ, which mainly 
contribute to E1(J8), are localized on opposite sides of the junction (see Fig. 
S2 in the Supporting Information). A charge transfer character is thus expected
for this excitation \cite{polymer-note}.

In order to better evaluate the charge transfer character of the 
excitations, we compute the spatial distribution of their \textit{hole} 
(\textit{h}) and \textit{electron} (\textit{e}) components \cite{baie+97ssc,ruin+02prl}. 
Taking advantage of the standard LCAO expansion of MOs \cite{jens07book} adopted here, 
\begin{equation}
\phi_{i}(\textbf{r}) = \sum_j a_{j i} \chi_j(\textbf{r}),
\label{lcao}
\end{equation}
where $\chi_j(\textbf{r})$ are Slater-type orbital basis functions and $a_{j i}$ are the projection coefficients, we express the \textit{probability density} of \textit{h} and \textit{e} for the $I^{th}$ excited state as:
\begin{equation}
\rho^{I}_{h}(\textbf{r}) = \sum_{\alpha \beta} \vert c_{\alpha \beta}^{I} \cdot \phi_{\alpha}(\textbf{r}) \vert^2
\label{rhoh}
\end{equation}
\begin{equation}
\rho^{I}_{e}(\textbf{r}) = \sum_{\alpha\beta} \vert c_{\alpha \beta}^{I} \cdot \phi_{\beta}(\textbf{r}) \vert^2,
\label{rhoe}
\end{equation}
where $c_{\alpha \beta}^{I}$ are CI coefficients weighting the contribution to the $I^{th}$ excitation of each transition $\alpha \rightarrow \beta$ from occupied ($\phi_{\alpha}$) to virtual ($\phi_{\beta}$) MOs. 
The spatial localization of each quasiparticle, \textit{h} and 
\textit{e}, can then be quantified by computing:
\begin{equation}
L_h = \frac{\sum_{y,z} \sum_{x\leq 0} \rho_h(x,y.z)}{\sum_{x,y,z} \rho_h(x,y.z)} \\
\label{loch}
\end{equation}
\begin{equation}
L_e = \frac{\sum_{y,z} \sum_{x\geq 0} \rho_e(x,y.z)}{\sum_{x,y,z} \rho_e(x,y.z)},
\label{loce}
\end{equation}
where the reference axes are centered at the flake, as shown in \ref{fig3}(a).
In the case of a homogeneous distribution, $L_{h/e}$ amounts to 50$\%$ of the charge density, while it is equal to 100$\%$ for a fully localized component.
For comparison, we also estimate the spatial localization of frontier orbitals, HOMO and LUMO, which mainly contribute to the first optically active excitation (see Supporting Information, Table SII).
The isosurfaces representing the \textit{hole} and \textit{electron} contributions 
to E1(J8) are shown in \ref{fig3}(b) and the corresponding values $L_h$ and $L_e$ 
are reported in the third column of \ref{table2}. 
We notice that, while the first excitation indeed presents a partial 
charge transfer character, the spatial localization of \textit{h} and 
\textit{e} is rather scarce. This feature, which is clearly inherited by 
frontier orbitals (see \ref{table2}), is also accompanied by a large oscillator 
strength of E1(J8), as a result of the non negligible overlap of the orbitals
contributing to the excitation.

\begin{table}[ht]
\begin{tabular*}{.58\textwidth}{c|ccc|ccc}
\hline  \hline
 & \multicolumn{3}{c|}{H//\ce{COCH3}} & \multicolumn{3}{c}{\ce{NH2}//F} \\
\hline
\textit{N}  & 6 & 7 & 8  & 6 & 7 & 8 \\
\hline \hline
$L_h$ & 67 $\%$ & 63 $\%$ & 56 $\%$ & 77 $\%$ & 87 $\%$ & 64 $\%$ \\
$L_e$ & 68 $\%$ & 75 $\%$ & 60 $\%$ & 87 $\%$ & 74 $\%$ & 65 $\%$ \\
\hline \hline
$L_{HOMO}$ & 80 $\%$ & 73 $\%$ & 58 $\%$ & 92 $\%$ & 96 $\%$ & 66 $\%$ \\
$L_{LUMO}$ &  79 $\%$  & 86 $\%$ & 62 $\%$ & 96 $\%$ & 92 $\%$ & 68 $\%$ \\
\hline \hline
\end{tabular*}
\caption{Spatial localization (per cent) of \textit{hole} ($L_h$) and \textit{electron} ($L_e$) components of the first optically active excitation of the H//\ce{COCH3} and \ce{NH2}//F GNJs, belonging to the families $N=6$, $N=7$ and $N=8$. The same for HOMO ($L_{HOMO}$) and LUMO ($L_{LUMO}$).}
\label{table2}
\end{table}

In order to enhance the charge transfer character of the excitation, we 
exploit the potential of edge functionalization, which allows
to maximize the energy offset at the interface by combining appropriate
functional groups according to the scheme summarized in \ref{fig1}.
Moreover, we investigate the generality of this mechanism by varying the 
junction width over the three families, i.e. $N=3p$, $N=3p+1$, and $N=3p+2$, 
which are known to present different electronic properties also upon 
functionalization \cite{cocc+11jpcc}.
In the following we consider prototypical junctions obtained by 
functionalizing one side with amino groups (\mbox{-\ce{NH2}}) and the other side 
with atomic fluorine (F).
The former have a very strong electron-donating character, as a
consequence of the lone pair of N, which is shared in the bond with
C edge atoms. The remarkable polarization of this covalent bond,
with about -0.5 $e^-$ on N and +0.1 $e^-$ on C -- as estimated through
Mulliken charge analysis -- induces a spreading of frontier orbitals
also onto N atoms. 
Noteworthy \mbox{-\ce{NH2}} termination is predicted to be quite stable and 
effective in modifying the electronic properties of quasi-1D graphene nanoribbons 
\cite{seit+10prb,cerv+07prb,ren+07condmat,kan+08jacs}.
On the other hand, F is strongly
electronegative, on account of its open shell electronic
configuration, typical of halogen atoms. 
Analogously to H, F atoms form $\sigma$ bonds at the edge: thanks to this similarity
F-termination has already been proposed to tune the electronic properties of graphene nanostructures \cite{zhen-dule08prb,gunl+10jpcl,zhu-su10jpcc}.

\begin{figure}
\centering
\includegraphics[width=.95\textwidth]{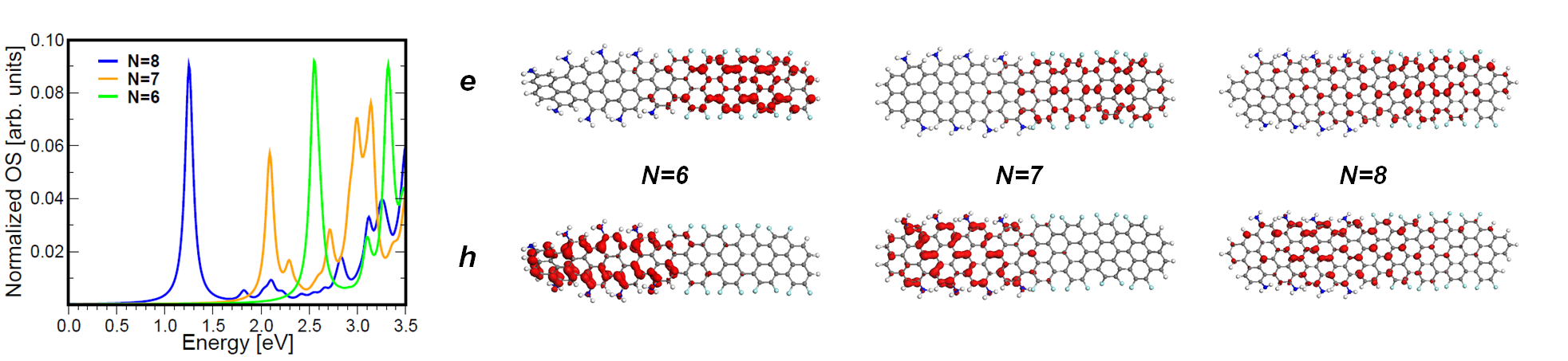}%
\caption{UV-Vis spectra of width modulated \ce{NH2}//F graphene nanojunctions, belonging to families $N=6$, $N=7$ and $N=8$. The curves represent oscillator strength (OS), normalized with respect to the number of C rings \textit{vs.} excitation energy and are convoluted with a Lorentzian function, with a broadening of 0.05 eV. Electron (\textit{e}) and hole (\textit{h}) probability densities are also shown for the first optically active excitations of each family.}
\label{fig4}
\end{figure}

We investigate the optical properties of \ce{NH2}//F nanojunctions of width parameters 
$N=6$, $N=7$ and $N=8$. 
The optical spectra and \textit{hole} and \textit{electron} densities for the first optically active excitation are displayed in \ref{fig4}; the spatial localizations are reported
in the last three columns of \ref{table2}.
For completeness, we also include in \ref{table2} 
the spatial localization for the corresponding excitation in H//\ce{COCH3} junctions; their probability densities are shown Figure S5 of the Supporting Information.
As can be seen from \ref{fig4}, the UV-Vis spectrum of $N=8$ \ce{NH2}//F 
junction retains the basic features already discussed for the H//\ce{COCH3} 
functionalization, but is red shifted by 0.25 eV as a consequence of the larger 
energy offset at the junction interface, which reduces the effective
optical gap. 
Consistently with the larger energy gap typical of the other two 
families, the first peak shows up instead at 2.09 and 2.55 eV for $N=7$ and 
$N=6$, respectively. The comparison of the three GNJs shows that the width 
parameter can be engineered to tune the energy of the first excitation, which 
is always bright and intense, over a wide range of about 1.5 eV.
The effectiveness of the new \ce{NH2}//F functionalization for the 
enhancement of the charge transfer character of the first optically active excitation is evident 
in the comparison of both probability densities (\ref{fig4}) and spatial 
localizations of \textit{electron} and \textit{hole} (\ref{table2}), the latter increasing up to 20$\%$ with respect 
to the H//\ce{COCH3} case \cite{dark-note}. Additional calculations performed on longer graphene 
nanojunctions (up to 5.5 nm) for the same \ce{NH2}//F functionalization indicate 
equal or further increased spatial localization, confirming the validity and 
generality of our results.
It is finally worth noting that in the $N=8$ case the spatial localization of 
\textit{hole} and \textit{electron} is again very similar to that of 
the frontier orbitals, which mainly contribute to the first excitation. 
This is not the case for $N=6$ and $N=7$, where HOMO $\rightarrow$ LUMO
transitions represent only the 20$\%$ of the total weight. Nonetheless, additional transitions 
involving higher-energy localized MOs contribute to the first excitation in the 
$N=6$ and $N=7$ GNJs, thus leading to a prominent charge transfer character of
the first excitation.

In summary, our results show that low energy excitations with
remarkable charge transfer character and large oscillator strength
can be obtained for edge-functionalized graphene nanojunctions. Both
prototypical H//\ce{COCH3} and \ce{NH2}//F nanojunctions have been
analyzed. A convenient choice of functional groups, which gives rise to opposite
gap region shifts relative to the H-terminated flake, helps maximizing the energy offset between
the two sides of the junction, thus enhancing the charge
transfer character of the first optically active excitation.
Furthermore, it is here demonstrated that  a wide 
energy range ($\sim$ 1.5 eV) can be spanned in the optical spectra by means of 
an appropriate width modulation, retaining at the same time the charge 
transfer character of the excitation.
The versatility of edge covalent functionalization, combined with the possibility 
of varying the width, makes these systems unique,
opening promising avenues for designing all-graphene optoelectronic nanodevices.

\begin{acknowledgement}
The authors are grateful to Enrico Benassi for stimulating and helpful discussions.
The authors acknowledge CINECA for computational support. This
work was partly supported by ``Fondazione Cassa di Risparmio di
Modena'', by the Italian Ministry of Foreign Affairs (Italy-USA
bilateral project) and by the Italian Ministry of University and
Research under FIRB grant ItalNanoNet. M.J.C. acknowledges support from
FAPESP and CNPq (Brazil).

\end{acknowledgement}
\begin{suppinfo}
We include the composition of the first excitations of homogeneously functionalized graphene nanoflakes and junctions discussed in the text.
The isosurfaces of occupied and virtual molecular orbitals close to HOMO and LUMO are displayed for the hydrogenated flake of width parameter $N=8$, as well as for all the considered junctions.
A perspective view of $N=8$ H//\ce{COCH3} and \ce{NH2}//F GNJs is presented, as well as the UV-Vis spectra and the \textit{hole} and \textit{electron} densities for the first optically active excitation of  H//\ce{COCH3} nanojunctions having width parameters $N=6$, $N=7$ and $N=8$.
\end{suppinfo}


\begin{mcitethebibliography}{54}
\providecommand*{\natexlab}[1]{#1}
\providecommand*{\mciteSetBstSublistMode}[1]{}
\providecommand*{\mciteSetBstMaxWidthForm}[2]{}
\providecommand*{\mciteBstWouldAddEndPuncttrue}
  {\def\EndOfBibitem{\unskip.}}
\providecommand*{\mciteBstWouldAddEndPunctfalse}
  {\let\EndOfBibitem\relax}
\providecommand*{\mciteSetBstMidEndSepPunct}[3]{}
\providecommand*{\mciteSetBstSublistLabelBeginEnd}[3]{}
\providecommand*{\EndOfBibitem}{}
\mciteSetBstSublistMode{f}
\mciteSetBstMaxWidthForm{subitem}{(\alph{mcitesubitemcount})}
\mciteSetBstSublistLabelBeginEnd{\mcitemaxwidthsubitemform\space}
{\relax}{\relax}


\bibitem{loh+10jmc}
Loh,~K.;\ \ Bao,~Q.;\ \ Ang,~P.;\ \ Yang,~J. \textit{J.~Mater.~Chem.~}
  \textbf{2010,} \textsl{20,} 2277--2289.

\bibitem{tyut+98jpcb}
Tyutyulkov,~N.;\ \ Madjarova,~G.;\ \ Dietz,~F.;\ \ Mullen,~K.
  \textit{J.~Phys.~Chem.~B} \textbf{1998,} \textsl{102,} 10183-10189.

\bibitem{wats+01cr}
Watson,~M.~D.;\ \ Fechtenkötter,~A.;\ \ Muellen,~K. \textit{Chem.~Rev.~}
  \textbf{2001,} \textsl{101,} 1267-1300.

\bibitem{wu+07cr}
Wu,~J.;\ \ Pisula,~W.;\ \ Muellen,~K. \textit{Chem.~Rev.~} \textbf{2007,}
  \textsl{107,} 718-747.

\bibitem{cai+10nat}
Cai,~J.;\ \ Ruffieux,~P.;\ \ Jaafar,~R.;\ \ Bieri,~M.;\ \ Braun,~T.;\ \
  Blankenburg,~S.;\ \ Muoth,~Matthiasand~Seitsonen,~A.~P.;\ \ Saleh,~M.;\ \
  Feng,~X.;\ \ Muellen,~K.;\ \ Fasel,~R. \textit{Nature~(London)~}
  \textbf{2010,} \textsl{466,} 470-473.

\bibitem{qian+08jacs}
Qian,~H.;\ \ Negri,~F.;\ \ Wang,~C.;\ \ Wang,~Z. \textit{J.~Am.~Chem.~Soc.~}
  \textbf{2008,} \textsl{130,} 17970.

\bibitem{wang+09sci}
Wang,~X.;\ \ Li,~X.;\ \ Zhang,~L.;\ \ Yoon,~Y.;\ \ Weber,~P.~K.;\ \ Wang,~H.;\
  \ Guo,~J.;\ \ Dai,~H. \textit{Science} \textbf{2009,} \textsl{324,} 768-771.

\bibitem{li+10jacs}
Li,~Y.;\ \ Gao,~J.;\ \ Di~Motta,~S.;\ \ Negri,~F.;\ \ Wang,~Z.
  \textit{J.~Am.~Chem.~Soc.~} \textbf{2010,} \textsl{132,} 4208.

\bibitem{shah+11}
Naghavi,~S.~S.;\ \ Gruhn,~T.;\ \ Alijani,~V.;\ \ Fecher,~G.~H.;\ \ Felser,~C.;\
  \ Medjanik,~K.;\ \ Kutnyakhov,~D.;\ \ Nepijko,~S.~A.;\ \ Schönhense,~G.;\ \
  Rieger,~R.;\ \ Baumgarten,~M.;\ \ Müllen,~K. \textit{J. Mol. Spectrosc.}
  \textbf{2011,} \textsl{265,} 95 - 101.

\bibitem{prez+08prb}
Prezzi,~D.;\ \ Varsano,~D.;\ \ Ruini,~A.;\ \ Marini,~A.;\ \ Molinari,~E.
  \textit{Phys.~Rev.~B} \textbf{2008,} \textsl{77,} 041404(R).

\bibitem{prez+07pssb}
Prezzi,~D.;\ \ Varsano,~D.;\ \ Ruini,~A.;\ \ Marini,~A.;\ \ Molinari,~E.
  \textit{Phys.~Status~Solidi~B} \textbf{2007,} \textsl{244,} 4124--4128.

\bibitem{prez+11tobe}
Prezzi, D.; Varsano, D.; Ruini, A.; Molinari, E. submitted \textbf{2011}.

\bibitem{yang+07nl}
Yang,~L.;\ \ Cohen,~M.;\ \ Louie,~S. \textit{Nano~Lett.~} \textbf{2007,}
  \textsl{7,} 3112-3115.

\bibitem{zhu-su10jpcc}
Zhu,~X.;\ \ Su,~H. \textit{J.~Phys.~Chem.~C} \textbf{2010,} \textsl{114,}
  17257-17262.

\bibitem{cocc+11jpcc}
Cocchi,~C.;\ \ Ruini,~A.;\ \ Prezzi,~D.;\ \ Caldas,~M.~J.;\ \ Molinari,~E.
  \textit{J.~Phys.~Chem.~C} \textbf{2011,} \textsl{115,} 2969-2973.

\bibitem{ridl-zern73tca}
Ridley,~J.;\ \ Zerner,~M. \textit{Theor.~Chem.~Acta} \textbf{1973,}
  \textsl{32,} 111-134.

\bibitem{cald+01apl}
Caldas,~M.~J.;\ \ Pettenati,~E.;\ \ Goldoni,~G.;\ \ Molinari,~E.
  \textit{Appl.~Phys.~Lett.~} \textbf{2001,} \textsl{79,} 2505--2507.

\bibitem{wang+04jacs}
Wang,~Z.;\ \ Tomovi\'{c},~Z.;\ \ Kastler,~M.;\ \ Pretsch,~R.;\ \ Negri,~F.;\ \
  Enkelmann,~V.;\ \ Muellen,~K. \textit{J.~Am.~Chem.~Soc.~} \textbf{2004,}
  \textsl{126,} 7794-7795.

\bibitem{dewa+85jacs}
Dewar,~M. J.~S.;\ \ Zoebish,~E.~G.;\ \ Healy,~E.~F.;\ \ Stewart,~J. J.~P.
  \textit{J.~Am.~Chem.~Soc.~} \textbf{1985,} \textsl{107,} 3902--3909.

\bibitem{MS-note}
AM1 and ZINDO/S calculations were performed using VAMP package included in
  Accelrys Materials Studio software, version 5.0
  (\url{http://accelrys.com/products/materials-studio}). We have chosen CI
  energy windows which include at least 4 eV below HOMO and 3 eV above LUMO, in
  order to ensure convergence.

\bibitem{son+06prl}
Son,~Y.-W.;\ \ Cohen,~M.~L.;\ \ Louie,~S.~G. \textit{Phys.~Rev.~Lett.~}
  \textbf{2006,} \textsl{97,} 216803.

\bibitem{baro+06nl}
Barone,~V.;\ \ Hod,~O.;\ \ Scuseria,~G.~E. \textit{Nano~Lett.~} \textbf{2006,}
  \textsl{6,} 2748--2754.

\bibitem{hod+08prb}
Hod,~O.;\ \ Barone,~V.;\ \ Scuseria,~G.~E. \textit{Phys.~Rev.~B} \textbf{2008,}
  \textsl{77,} 035411.
  
\bibitem{zhan+05ol}
Zhang,~Q.;\ \ Prins,~P.;\ \ Jones,~S.~C.;\ \ Barlow,~S.;\ \ Kondo,~T.;\ \
  An,~Z.;\ \ Siebbeles,~L. D.~A.;\ \ Marder,~S.~R. \textit{Org.~Lett.}
  \textbf{2005,} \textsl{7,} 5019-5022.

\bibitem{kiku+07ol}
Kikuzawa,~Y.;\ \ Mori,~T.;\ \ Takeuchi,~H. \textit{Org.~Lett.} \textbf{2007,}
  \textsl{9,} 4817-4820.

\bibitem{dist-note}
The impact of distortions on the optical properties of both graphene nanoflakes and junctions 
  is found to be negligible. To show this, we have taken the backbone of a 
  fully optimized \ce{NH2}-graphene flake, where distortions are maximized,
  we have removed functionalization and then H-saturated the dangling bonds
  without any further geometrical optimization.
  The optical properties of this system compared to those of a fully
  optimized hydrogenated flake -- completely flat and undistorted --
  do not show any significant modification, except for a small rigid shift of the spectra.
  Further details are reported in the Supporting Information.

\bibitem{polymer-note}
An analogous behavior is observed also in conjugated copolymers, characterized
  by an alternation of donor and acceptor units, which induce a localization of
  frontier orbitals and consequently of \textit{hole} and \textit{electron}
  densities.
  See e.g. Jespersen,~K.~G.;\ \ Beenken,~W. J.~D.;\ \ Zaushitsyn,~Y.;\ \ Yartsev,~A.;\ \
  Andersson,~M.;\ \ Pullerits,~T.;\ \ Sundstrom,~V. \textit{J.~Chem.~Phys.~}
  \textbf{2004,} \textsl{121,} 12613-12617.

\bibitem{baie+97ssc}
Baierle,~R.~J.;\ \ Caldas,~J.;\ \ Molinari,~E.;\ \ Ossicini,~S. \textit{Solid
  State Commun.~} \textbf{1997,} \textsl{102,} 545-549.

\bibitem{ruin+02prl}
Ruini,~A.;\ \ Caldas,~M.~J.;\ \ Bussi,~G.;\ \ Molinari,~E.
  \textit{Phys.~Rev.~Lett.~} \textbf{2002,} \textsl{88,} 206403.

\bibitem{jens07book}
Jensen,~F. \textit{Introduction to Computational Chemistry;} Wiley: Second ed.;
  2007.

\bibitem{seit+10prb}
Seitsonen,~A.~P.;\ \ Saitta,~A.~M.;\ \ Wassmann,~T.;\ \ Lazzeri,~M.;\ \
  Mauri,~F. \textit{Phys.~Rev.~B} \textbf{2010,} \textsl{82,} 115425.

\bibitem{cerv+07prb}
Cervantes-Sodi,~F.;\ \ Cs\'anyi,~G.;\ \ Piscanec,~S.;\ \ Ferrari,~A.~C.
  \textit{Phys.~Rev.~B} \textbf{2008,} \textsl{77,} 165427.

\bibitem{ren+07condmat}
Ren,~H.;\ \ Li,~Q.;\ \ Su,~H.;\ \ Shi,~Q.~W.;\ \ Chen,~J.;\ \ Yang,~J.
  \textit{Cond-Mat} \textbf{2007,}  arXiv0711.1700v1.

\bibitem{kan+08jacs}
Kan,~E.-j.;\ \ Li,~Z.;\ \ Yang,~J.;\ \ Hou,~J.~G. \textit{J.~Am.~Chem.~Soc.~}
  \textbf{2008,} \textsl{130,} 4224-4225.

\bibitem{zhen-dule08prb}
Zheng,~H.;\ \ Duley,~W. \textit{Phys.~Rev.~B} \textbf{2008,} \textsl{78,}
  045421.

\bibitem{gunl+10jpcl}
Gunlycke,~D.;\ \ Mintmire,~J.~W.;\ \ White,~C.~T. \textit{J. Phys. Chem. Lett.}
  \textbf{2010,} \textsl{1,} 1082.

\bibitem{dark-note}
In the case of $N=7$ H//\ce{COCH3} nanojunction, the charge transfer excitation,
  whose results for the \textit{hole} and \textit{electron} components are reported in \ref{table2},
  is actually the second excitation. This is due to the electronic properties of $N=7$ graphene
  flakes, which present a dark excitation at lowest energy.
  This feature disappears upon \ce{NH2}//F functionalization, where the first excitation
  is optically active and presents charge transfer character for all the considered \textit{families}.
  See Supporting Information for more detailed discussion.
  
\end{mcitethebibliography}
\providecommand*{\mcitethebibliography}{\thebibliography}
\csname @ifundefined\endcsname{endmcitethebibliography}
{\let\endmcitethebibliography\endthebibliography}{}


\end{document}